\documentclass[12pt]{article}
\date{}
\begin{document}

{\bf Event-independence, collective-independence,EPR-Bohm }

{\bf experiment and incompleteness of quantum mechanics}

\medskip

Andrei Khrennikov

\medskip

{\it International Center for Mathematical
Modeling}

{\it in Physics and Cognitive Sciences,}

{\it  University of V\"axj\"o, S-35195, Sweden}\footnote{Email:Andrei.Khrennikov@msi.vxu.se}

\medskip

\begin{abstract} We analyse notion of independence in the 
EPR-Bohm framework by using comparative analysis of independence
in conventional and frequency probability theories. Such an
analysis is important to demonstrate that Bell's inequality
was obtained by using totally unjustified assumptions (e.g. 
the  Bell-Clauser-Horne factorability condition). Our frequency analysis
also demonstrated that Gill-Weihs-Zeilinger-Zukowski's
arguments based on "the experimenter's
freedom to choose settings" to support the standard Bell
approach are neither justified by the structure of the EPR-Bohm
experiment. Finally, our analysis supports the original 
Einstein's viewpoint that quantum mechanics is simply not complete.
\end{abstract}

Preprint of R. Gill, G. Weihs, A. Zeilinger, M. Zukowski
[1] stimulated the interest to the role of independence conditions
in the EPR-Bohm framework. I recall that preprint [1] was published
as the rigid critical reply to works of K. Hess and W. Philipp, see
e.g. [2]. Here I do not consider Hess-Philipp arguments, but only 
their conclusion: in general we do not have Bell's inequality, since
the  Bell-Clauser-Horne factorability condition (sometimes called locality condition):
\begin{equation}
\label{BF}
{\bf P} (A,B^\prime/ a,b^\prime, \lambda)={\bf P} (A/ a, \lambda){\bf P} (B^\prime/b^\prime, \lambda)
\end{equation}
can be violated in very natural models.

The main Gill-Weihs-Zeilinger-Zukowski's counter-argument against  Hess-Philipp's
model is the following one: "However, they themselves have neglected the experimenter's
freedom to choose settings..." In fact, this is independence condition. Therefore
it would be useful to provide general analysis of the role of independence conditions
in the  EPR-Bohm experiment. This analysis is deeply related to the very foundations
of probability theory, namely Kolmogorov (measure-theoretical model, i.e. conventional
model) and von Mises (frequency model, nonconventional, but very experimental).
Some parts of such an analysis can be found in my preprints [3], [4] (the last one 
also contains critical analysis of Gill-Weihs-Zeilinger-Zukowski's as well as 
Hess-Philipp's arguments). Here we concentrate us to the role of independence 
condition.
 
Our frequency analysis demonstrates that the  Bell-Clauser-Horne factorability condition
should be violated simply due to dependence of corresponding collectives (random sequences)
of hidden variables, HV. So there is no Bell inequality. Collective dependence
is a consequence of original EPR-correlations in pairs of particles. Therefore,
as it was originally claimed by Einstein, Podolsky and Rosen, [5], quantum mechanics
is not complete. "The experimenter's
freedom to choose settings" that was the cornerstone of Gill-Weihs-Zeilinger-Zukowski's 
considerations does not change anything: corresponding collectives are still dependent.

In the {\bf  frequency approach} (von Mises, [6]) probability is defined as the limit of relative
frequencies in a collective (random sequence) produced by some experimental device.
Two collective are said to be {\it independent} if corresponding
experimental devices works independently. This implies factorization of probabilities
in the multi-collective created by {\it combination} of two independent collectives,
see [6], [7] for the details.

In the {\bf measure-theoretical approach} (Kolmogorov, [8]) probability is defined 
as an abstract (normalized) measure. Two events are said to be independent if 
the probability of their intersection is factorized.

Gill, Weihs, Zeilinger, Zukowski (as well as all others)  
use Kolmogorov independence - event independence and I shall use
von Mises independence: collective independence. These are two very different
notions. The reader  can find in von Mises book [6] as well as in my book [7]
examples in that factorization of probability can occur simply as the result of the play with numbers. 
This is event independence. In the opposite to event independence,   
collective independence is the physical notion.
And the EPR experiment evidently demonstrated that collective independence
is the right notion to describe real physical experiments.

We consider the general probabilistic scheme of the EPR-Bohm experiment.

There are two physical systems, $U_{\rm{left}}$ and $U_{\rm{right}},$
producing the settings of measurement devices, $a$
and $b^\prime,$ and there is third system, $U$, producing correlated particles. 
I want to underline
(and the frequency analysis immediately shows this) that, despite the
independence of $U_{\rm{left}}$,  $U_{\rm{right}}$ ,
 and $U$, collectives produced by pairs ($U_{\rm{left}}$,  $U$) and
($U_{\rm{right}}$, $U$) need not be independent. Moreover, they should be
dependent due to the presence of the common physical system, namely $U.$
So we can do what ever we want with $U_{\rm{left}}$,  $U_{\rm{right}}$ and even 
with $U.$ But the whole statistical structure of the experiment induces
dependence of collectives.

In our private Email discussion R. Gill noticed:
"But note, I am talking about independence between
the physical system generating the settings (e.g., 
tossing coins), and the physical system which could
accept either setting and then output the outcomes.

This is a different independence from the one you
were talking about. I would say, that it is a reasonable
physical assumption that physically independent
subsystems of the world (a coin toss here, some photons
there) exist. If you deny this then of course anything
is possible. Do you deny the possibility of tossing
a coin independently of sending a photon through a
polarizer?"

I do not deny this possibility. Frequency analysis
demonstrated that the main problem is the presence 
of the common physical system $U$ (producing correlated
particles) in the left-hand side as well as the right-hand
side collectives. Thus the freedom of experimenters
playing with devices $U_{\rm{left}}$ and $U_{\rm{right}}$ 
does not destroy $U$-dependence.

We present the formal collective description of the model.

Let $\lambda_j, j=1,2,...$ be the value of the HV for the $j$th pair of correlated particles
$(\pi^1_j, \pi^2_j)$ produced at the instance of time $t_j=j.$
We consider two sequences of pairs and a sequence of triples (three collectives):
$$
x_{\omega_{\rm{left}}, \lambda}= \{ (\omega_{\rm{left}\; 1}, \lambda_1),...., (\omega_{\rm{left} \;N}, \lambda_N),...\}\;,
$$
$$
x_{\omega_{\rm{right}}, \lambda}= \{ (\omega_{\rm{right} \;1}, \lambda_1),...., (\omega_{\rm{right}\; N}, \lambda_N),...\}\;,
$$
and
$$
x_{\omega_{\rm{left}}, \lambda, \omega_{\rm{right}}}= \{ (\omega_{\rm{left}\; 1}, \lambda_1, \omega_{\rm{right} \;1}),...., 
(\omega_{\rm{left} \;N}, \lambda_N, \omega_{\rm{right}\; N}),...\}\;,
$$
where $\omega_{\rm{left} \;j}$ and $\omega_{\rm{right}\; j}$  
are internal states of apparatuses $U_{\rm{left}}$ and $U_{\rm{right}},$ 
respectively.\footnote{ In this framework it is not important that the experimenters
have the freedom of choice of experimental setting for devices 
$U_{\rm{left}}$ and $U_{\rm{right}}.$ }
Due to the presence of the common parameter $\lambda$ in both collectives
$x_{\omega_{\rm{left}}, \lambda}$ and $x_{\omega_{\rm{right}}, \lambda}$, they are not
independent. Therefore the probability
distribution of the collective $x_{\omega_{\rm{left}}, \lambda, \omega_{\rm{right}}}$
could not be factorized.

As we have already remarked, in fact, our frequency analysis gives strong probabilistic support to 
the original EPR-arguments. It seems that A. Einstein rightly pointed to incompleteness
of quantum mechanics. This incompleteness is a consequence of the impossibility to describe
the correlation HV $\lambda$ in quantum formalism. Nevertheless, quantum formalism
gives the right answer in the EPR-framework, since dependence of collectives is 
coded into corresponding quantum state. However, even by having the right answer 
in quantum formalism we could not explain the origin of this answer. This induces 
unusual explanations such as e.g. {\it nonlocality.}\footnote{ We also notice that in our frequency analysis it is really
impossible to find even a trace of some nonlocalty that is typically related 
to the EPR-experiment.}  We remark that our paper could not be used as an argument 
against nonlocalty. Quantum mechanics could be nonlocal. However, Bell's 
inequality definitely could not be used as an argument in nonlocality story.

Finally, we remark that if there is no independence, in particular,
the  Bell-Clauser-Horne factorability condition, then we should
get modified Bell's inequalities, see [9]-[11], instead of the original
Bell-type inequalities. Inequalities that I have obtained in [9], [10]
need not be violated in quantum theory.

I would like to thank R. Gill for intensive discussions that clarified
my own views to the EPR-experiments.

{\bf REFERENCES}

1. R. Gill, G. Weihs, A. Zeilinger, M. Zukowski, {\it Comment on "Exclusion of time in the theorem 
of Bell" by K. Hess and W. Philipp.} quant-ph/0204169 (2002).

2. K. Hess and W. Philipp, {\it Einstein-separability, time related hidden 
parameters for correlated spins, and the theorem of Bell.} quant-ph/0103028;
{\it Proc. Nat. Acad. Sci. USA,} {\bf 98}, 14224 (2001); 
{\it Proc. Nat. Acad. Sci. USA,} {\bf 98}, 14228 (2001); 
{\it Europhys. Lett.}, {\bf 57}, 775 (2002).

3. A. Khrennikov, {\it Einstein and Bell, von Mises and Kolmogorov: reality,
locality, frequency and probability.} quant-ph/0006016 (2000).

4. A. Khrennikov, {\it Comment on Hess-Philipp anti-Bell and 
Gill-Weihs-Zeilinger-Zukowski anti-Hess-Philipp
arguments.} quant-ph/0205022.

5. A. Einstein, B. Podolsky, N. Rosen,  Phys. Rev., {\bf 47}, 777--780
(1935).

6. R.  von Mises, {\it The mathematical theory of probability and
 statistics}. Academic, London (1964).

7. A. Khrennikov, {\it Interpretations of Probability.}
VSP Int. Sc. Publishers, Utrecht (1999).

8.  A. N. Kolmogoroff, {\it Grundbegriffe der Wahrscheinlichkeitsrechnung.}
Springer Verlag, Berlin (1933); reprinted:
{\it Foundations of the Probability Theory}. 
Chelsea Publ. Comp., New York (1956).

9. A. Yu. Khrennikov, A perturbation of CHSH inequality induced by fluctuations 
of ensemble distributions. {\it J. of Math. Physics}, {\bf 41}, N.9, 5934-5944, (2000).

10.  A. Yu. Khrennikov, Non-Kolmogorov probability models and modified Bell's
inequality. {\it J. of Math. Physics,} {\bf 41}, N.4, 1768-1777 (2000).

11. O. G. Smolyanov, A. Truman, {\it On Bell-Khrennikov inequalities.} Preprint of Moscow
State University, March-2002.

\end{document}